\documentclass[prl,amsfonts,amssymb,amsmath,floats,twocolumn,aps]{revtex4}
\usepackage[pdftex]{graphicx}
\usepackage{dcolumn}
\usepackage{bm}
\usepackage{color}

\usepackage[pdftex,colorlinks=true,linkcolor=blue,citecolor=blue,filecolor=blue]{hyperref}

\newcommand{\ff}[1]{{\boldsymbol #1}}
\newcommand{\ca}[1]{{\cal #1}}

\newcommand{\bi}{\begin{itemize}}
\newcommand{\ei}{\end{itemize}}
\newcommand{\be}{\begin{equation}}
\newcommand{\ee}{\end{equation}}
\newcommand{\ba}{\begin{eqnarray}}
\newcommand{\ea}{\end{eqnarray}}

\newcommand{\parag}[1]{\paragraph{\color{blue}#1}}

\begin{document} 

\title{
Anomalous spin precession under a geometrical torque
}

\author{Christopher Stahl and Michael Potthoff}
\affiliation{I. Institute for Theoretical Physics, Universit\"at Hamburg, Jungiusstra\ss e 9, 20355 Hamburg, Germany}
\affiliation{The Hamburg Centre for Ultrafast Imaging, Luruper Chaussee 149, 22761 Hamburg, Germany}

\begin{abstract}
Precession and relaxation predominantly characterize the real-time dynamics of a spin driven by a magnetic field and coupled to a large Fermi sea of conduction electrons.
We demonstrate an anomalous precession with frequency higher than the Larmor frequency or with inverted orientation in the limit where the electronic motion adiabatically follows the spin dynamics. 
For a classical spin, the effect is traced back to a geometrical torque resulting from a finite spin Berry curvature. 
\end{abstract} 
 
\maketitle 
  
%----------------------------------------------------------------------------------------------------------
\parag{Introduction.}

Topological concepts are more and more recognized as organizing principles for condensed-matter systems.
Celebrated examples are new types of phase transitions in two dimensions involving topological defects \cite{Kos74,KT72},
the explanation of the quantization of the Hall conductance in two-dimensional electron gases \cite{TKNN82}, or Haldane's conjecture for one-dimensional spin chains \cite{Hal83a,Hal83b}. 
In particular, topological invariants based on the Berry phase \cite{Ber84} play an important role for electronic properties of solids \cite{XCN10,HK10} and in magnetism \cite{Bru05}. 

Via the adiabatic theorem \cite{Mes61}, the Berry phase comes into play in quantum systems with slowly varying external parameters. 
More generally, the Berry phase, or actually the Berry curvature, plays an ubiquitious role in any quantum-classical hybrid system with slow classical and fast quantum degrees of freedom \cite{KI85}. 
Integrating out the fast degrees of freedom, generates a geometrical {\em force} in the slow classical subsystem \cite{ZW06}. 
This has been a major topic in semiclassical electron dynamics in crystals \cite{Res00}.
In the field of magnetism, pioneering work \cite{WZ88,NK98,NWK+99} on adiabatic long-wavelength magnon dynamics has put the focus on the 
spin Berry curvature which enters the linearized equation of motion. 

%----------------------------------------------------------------------------------------------------------
\parag{Geometrical torque in spin dynamics.}

Here, we report a striking effect of a geometrical {\em torque} which strongly renormalizes the precession frequency of a classical spin coupled to conduction electrons. 
The torque is caused by the spin Berry curvature of the electrons \cite{note}.
The resulting anomalous precession frequency can be studied within the $s$-$d$ exchange (Vonsovsky-Zener) model \cite{VZ} where the spin is taken as a classical dynamical variable.
This model comprises, in a nutshell, the main concepts of microscopic spin dynamics \cite{TKS08,SHNE08,BMS09,EFC+14}, i.e., spin precession, Gilbert damping and relaxation, nutation etc.\ \cite{SP15}.

Concretely, we find that a classical spin of length $S$ driven by an external magnetic field $B$ and exchange coupled to a Fermi sea of conduction electrons precesses with a frequency $\omega_{\rm p}$ which may dramatically differ from the standard Larmor frequency $\omega_{\rm L} \propto B$.
This is due to a geometrical torque which adds to the torque exerted on the spin by $B$ and which stems from the spin Berry curvature $\Omega$ of the conduction electrons.
$\Omega$ is basically independent of the details of the electronic structure. 
It is nonzero, e.g., for a nonmagnetic Fermi sea with an odd number of electrons. 
In this case, and assuming perfect adiabaticity, an entirely analytical computation yields
\be
\omega_{\rm p} = \frac{\omega_{\rm L}}{1 - s / S} \: ,
\label{eq:omega}
\ee
where $s=1/2$ is the electron spin quantum number. 
For a classical-spin length $S>s$ this predicts a faster precession; for $S<s$ the orientation would be inverted.

Magnetic atoms on nonmagnetic surfaces, the magnetic state of which is controlled by a polarized STM tip or by the proximity to switchable magnetic islands, represent a realistic example \cite{Wie09}. 
However, as the precession frequency has a direct impact on several observables \cite{TKS08,SHNE08,BMS09,EFC+14}, we expect that geometrical torques play an important role in spin dynamics {\em generally}.
This is opposed to coupled systems of (fast) electrons and (slow) nuclei where the Berry phase can be viewed \cite{MAKG14} as an artifact of the Born-Oppenheimer approximation.

Anomalous precession is in fact found in large regions of the parameter space, also {\em beyond} the adiabatic limit, as verified here by numerical calculations for the full $s$-$d$ exchange model.
In addition, the quantum-spin (Kondo) model \cite{Kondo} is considered, to demonstrate robustness against quantum fluctuations. 
On the other hand, spin-only theories, such as approaches based on the Landau-Lifschitz-Gilbert (LLG) equation \cite{llg}, do not include the geometrical torque. 

%----------------------------------------------------------------------------------------------------------
\parag{Model.}

\begin{figure}
\includegraphics[width=0.98\columnwidth]{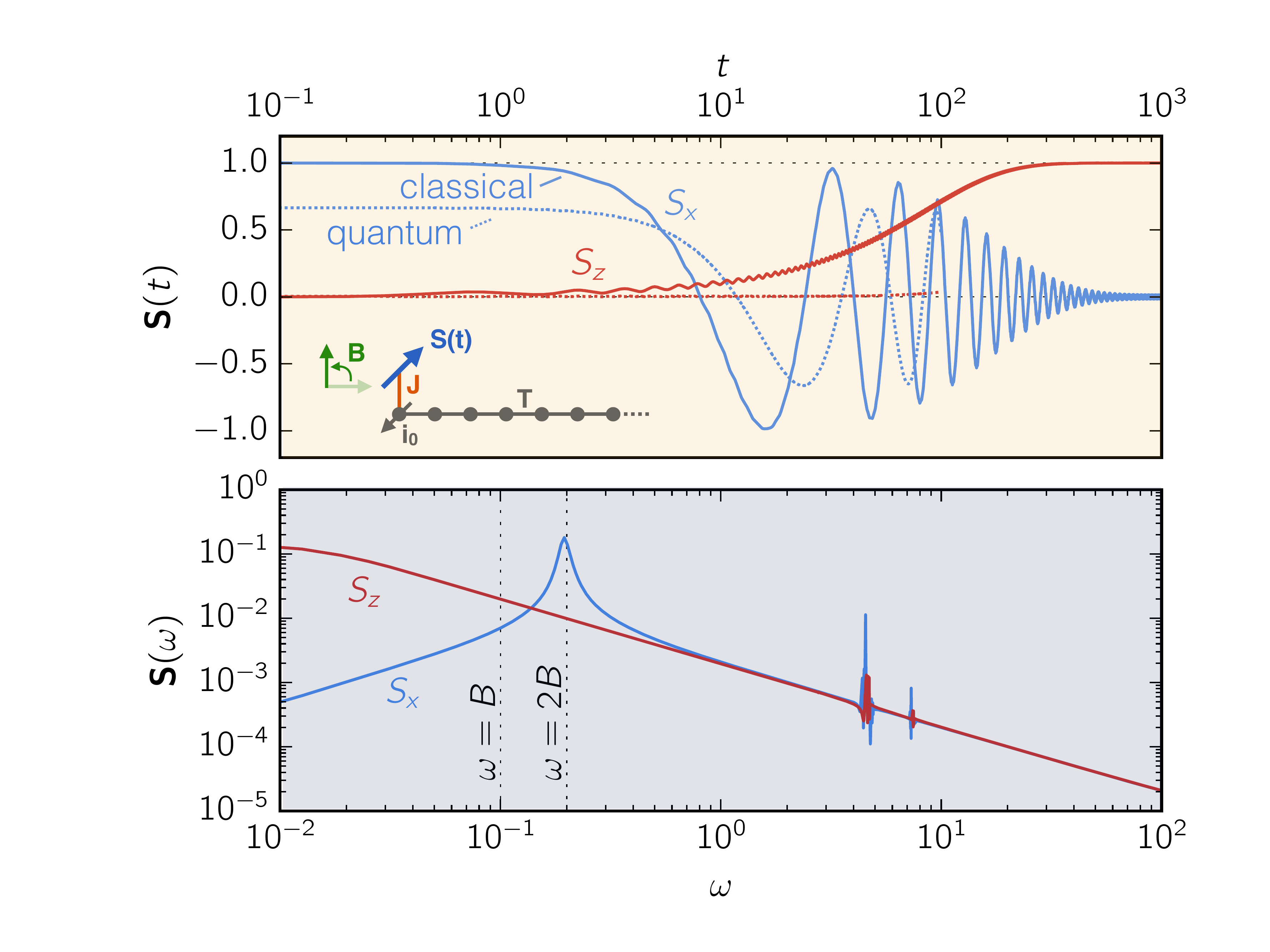}
\caption{
{\em Top:} Time dependence of the $x$- and $z$-component of the classical spin from the
numerical solution of the model (\ref{eq:ham}) on an open chain with $L=1001$ sites. 
The spin with $| \ff S |=S=1$ couples to site $i_{0}=1$ with $J=10$.
Initially, at $t=0$, the system is in the ground state and $\ff S = S \hat{x}$. 
For $t>0$, a field ${\ff B}=0.1 \hat{z}$ drives the spin.
The n.n.\ hopping $T=1$ (and $\hbar \equiv 1$) fixes the energy (and time) scale.
Dotted lines: the same but for a quantum spin $S=1$ using t-DMRG ($L=51, J=10, B=0.1$).
{\em Bottom:} Fourier transform.
}
\label{fig:soft}
\end{figure}

We consider a classical spin $\ff S$ of length $|\ff S| = S$, which is coupled to a system of $N$ itinerant and noninteracting conduction electrons (see inset in Fig.\ \ref{fig:soft}).
The electrons hop with amplitude $-T$ between non-degenerate orbitals on nearest-neighboring sites of a $D$-dimensional lattice with $L$ sites.
$n=N/L$ is the average conduction-electron density which we choose as $n=1$ (half filling). 
Furthermore, we set $T \equiv 1$ to fix energy and time units ($\hbar \equiv 1$).
The coupling is a local antiferromagnetic exchange of strength $J>0$ between $\ff S$ and the local quantum spin $\ff s_{i} = \frac{1}{2} \sum_{\sigma \sigma'} c^{\dagger}_{i\sigma} \ff \tau_{\sigma\sigma'} c_{i\sigma'}$ at the site $i=i_{0}$ ($\ff \tau$ is the vector of Pauli matrices, and $\sigma=\uparrow, \downarrow$).
The dynamics of this quantum-classical hybrid system \cite{Elz12} is determined by the Hamiltonian
\be
{H} = - T \sum_{\langle i,j \rangle, \sigma} c^{\dagger}_{i\sigma} c_{j\sigma} 
+ 
J \ff s_{i_0} \ff S 
-
\ff B \ff S 
\: .
\label{eq:ham}
\ee
This is the famous $s$-$d$ exchange model \cite{VZ}.
Here, $c_{i\sigma}$ annihilates an electron at site $i=1,...,L$ with spin projection $\sigma$, and $\ff B$ is an external magnetic field which drives the classical spin ($g$-factor and Bohr magneton have been absorbed in the definition of $B$). 
If $\ff S$ was a quantum spin with $S=1/2$, Eq.\ (\ref{eq:ham}) would represent the single-impurity Kondo model \cite{Kondo}.

%----------------------------------------------------------------------------------------------------------
\parag{Spin dynamics.}

Real-time dynamics will be initiated by a sudden change of the field direction from, say, $\hat{x}$ to $\hat{z}$ direction.
We assume that initially, at time $t=0$, the system is in its ground state, i.e., the classical spin is aligned to the external field $\ff S(t=0) \propto \hat{x}$, and the conduction electrons occupy the lowest one-particle eigenstates of the non-interacting ($J=0$) Hamiltonian for the given $\ff S(t=0)$.

For $J=0$, the spin 
will perform an undamped precessional motion around $\hat{z}$ with Larmor frequency $\omega_{\rm p}=B$, according to the Landau-Lifschitz (LL) equation $\dot{\ff S} = \ff S \times \ff B$.
Damping and finally relaxation $\ff S(t) \to S \hat{z}$ is phenomenologically described by the LLG equation \cite{llg} and is also found microscopically from the model (\ref{eq:ham}) for $J>0$ \cite{TKS08,BMS09,SP15,SRP16a}.
Besides precession and damping, the electronic system causes a weak nutation of the spin \cite{But06,WC12,SRP16b}. 
While damping $\propto \dot{\ff S}$ and nutation $\propto \ddot{\ff S}$ are higher-order effects \cite{BNF12}, spin precession is the most elementary phenomenon in this context. 
As we will show, however, the precession frequency crucially depends on the presence of a geometrical torque caused by the conduction-electron system.
This is beyond an LLG-type approach.

%----------------------------------------------------------------------------------------------------------
\parag{Equations of motion.}

The model (\ref{eq:ham}) can be solved numerically for large lattices ($\sim 1000$ sites) by applying a high-order Runge-Kutta technique to a nonlinear system of ordinary differential equations:
The equation of motion for $\ff S(t)$ is derived as the canonical equation from the Hamilton function $H_{\rm class} = \langle H \rangle$ (see Refs.\ \onlinecite{Hes85,Hal08,Elz12} and references therein for a general discussion).
We find an LL-type equation 
\be
\frac{d}{dt} \ff S(t)
=
J \langle \ff s_{i_0} \rangle_{t} \times \ff S(t)
-
\ff B(t) \times \ff S(t) \: ,
\label{eq:eoms}
\ee
where the additional time-dependent internal Weiss field $J \langle \ff s_{i_0} \rangle_{t}$ is obtained via
$
  \langle \ff s_{i_0} \rangle_{t} 
  =
  \frac{1}{2} \sum_{\sigma\sigma'} \rho_{i_{0} \sigma, i_{0}\sigma'}(t) \, \ff \tau_{\sigma'\sigma} 
%\label{eq:si0}  
$,
from the one-particle reduced density matrix 
$
\rho_{ii',\sigma\sigma'}(t) \equiv \langle c_{i'\sigma'}^{\dagger} c_{i\sigma} \rangle_{t} 
%\label{eq:densitymatrix}
$,
which is an expectation value in the conduction-electrons' state $|\Psi(t)\rangle$ at time $t$.
$\ff \rho(t)$ obeys a von Neumann equation of motion
\be
  i \frac{d}{dt} \ff \rho(t) = [ \ff T(t) , \ff \rho(t) ] \: , 
\label{eq:rhoeom}
\ee 
where the matrix $\ff T(t)$ with elements
$
  T_{i\sigma,i'\sigma'}(t) = -T \delta_{\langle ij \rangle} \delta_{\sigma\sigma'} + \delta_{ii_{0}} \delta_{i'i_{0}}
  \frac{J}{2} (\ff S(t) \ff \tau)_{\sigma\sigma'}
%\label{eq:hopp}
$
consists of the physical hopping and the contribution of the Weiss field produced by the spin which couples to the site $i_{0}$.
Here, $\delta_{\langle ij \rangle} =1$ if $i, j$ are nearest neighbors and zero else.

%%%%%%%%%%%%%%%%%%%%%%%%%%%%%%%%%%%%%%%%%%%%%%%%%%%%%%%%%%%%%%
%%%%%%%%%%%%%%%%%%%%%%%%%%%%%%%%%%%%%%%%%%%%%%%%%%%%%%%%%%%%%%

%----------------------------------------------------------------------------------------------------------
\parag{Anomalous precession frequency.}

Fig.\ \ref{fig:soft} presents numerical results for a spin of length $S=1$. 
After the sudden switch of the field direction, the spin exhibits a precessional motion around $\hat{z}$, see the oscillations of its $x$ component, and relaxes to the new field direction on a time scale of $\tau \sim 10^{3}$. 
Superimposed to the damping are tiny oscillations of the $z$ component which have been attributed \cite{SRP16b} to nutation. 
We will argue below that with a strong exchange coupling $J=10$ and a weak field $B=0.1$ (as compared to $T=1$), as chosen in this example, we are in parameter regime where the adiabatic approximation is nearly perfectly justified. 

Hence, Eq.\ (\ref{eq:omega}) applies and predicts a doubling of the precession frequency: $\omega_{\rm p} \approx 2 B$.
This is clearly visible in the lower panel of Fig.\ \ref{fig:soft} which shows that the Fourier transform $S_{x}(\omega)$ is strongly peaked at $2B$.
Weaker peaks at high frequencies correspond to the nutational motion.

%----------------------------------------------------------------------------------------------------------
\parag{Adiabatic spin dynamics.}
 
We now show that the anomalous precession frequency is the result of a geometrical torque.
If the spin dynamics is slow as compared to the typical time scale of the electron dynamics, it is reasonable to assume that the conduction-electron state at time $t$ is the ground state for the given spin orientation $\ff S(t)$ at time $t$:
\be
  | \Psi(t) \rangle = | \Psi_{0}[\ff S(t)] \rangle \: . 
\label{eq:cons}
\ee
To exploit this as a holonomic constraint of the coupled spin-electron dynamics and to eliminate the electron degrees of freedom, we first rederive the equations of motion from the Lagrangian
\be
  \ca L
  =
  \ff A(\ff S) \dot{\ff S} + \langle \Psi | i \partial_{t} | \Psi \rangle - \langle \Psi | H | \Psi \rangle \: .
\label{eq:lagrangian}
\ee
Here, $\ff A(\ff S)$ is conjugate to $\ff S$ and must satisfy $\nabla_{\ff S} \times \ff A(\ff S) = -\ff S$ on the Bloch sphere $|\ff S| = S$ with radius $S$. 
This can be found from the solution of the Dirac magnetic monopole problem \cite{Dir60} by interpreting $\ff A$ as the magnetic vector potential and $\ff S$ as the position vector. 
The constraints $|\ff S| = S$ and $\langle \Psi | \Psi \rangle=1$ can be accounted for using appropriate Lagrange multipliers.
It is straightforward to verify that $\ca L$ generates the equation of motion (\ref{eq:eoms}) and Schr\"odinger's equation for $| \Psi (t) \rangle$ or, equivalently, Eq.\ (\ref{eq:rhoeom}).

Within the adiabatic approximation and for a single spin, we get the effective spin-only Lagrangian from Eq.\ (\ref{eq:lagrangian}) by eliminating the electronic degrees of freedom using Eq.\ (\ref{eq:cons}).
The resulting equation of motion is:
\be
\dot{\ff S} 
= 
J \langle \ff s_{i_0} \rangle^{(0)} \times \ff S
-
\ff B \times \ff S
+
\ff S \times (\ff \Omega(\ff S) \times \dot{\ff S})
\: ,
\label{eq:eomberry}
\ee
with an implicit $t$-dependence of $\langle \ff s_{i_0} \rangle^{(0)} \equiv \langle \Psi_{0}[\ff S(t)]  | \ff s_{i_0} | \Psi_{0}[\ff S(t)] \rangle$ via the instantaneous ground state $| \Psi_{0}[\ff S(t)] \rangle$.
The last term represents the geometrical torque and involves the (implicitly $t$-dependent) pseudovector $\ff \Omega = \ff \Omega[\ff S(t)]$ with components $\Omega_{\alpha} = - \sum_{\beta\gamma} \epsilon_{\alpha\beta\gamma} \Omega_{\beta\gamma} / 2$ ($\alpha=x,y,z$) formed by the three independent elements of the spin Berry curvature
\be
  \Omega_{\alpha\beta}(\ff S)
  =
  \frac{\partial}{\partial S_{\alpha}}
  \langle \Psi_{0}[\ff S] \big| 
  i \frac{\partial}{\partial S_{\beta}}
   \big| \Psi_{0}[\ff S] \rangle
   -
   (\alpha \leftrightarrow \beta)
\label{eq:curv}
\ee
which is a rank-2 antisymmetric tensor. 

Simplifications of the adiabatic equation of motion (\ref{eq:eomberry}) arise from symmetries: 
The classical spin essentially acts as an external local magnetic field breaking the SU(2) spin rotational symmetry of the conduction-electron system, but the remaining rotational symmetry around the axis given by the radial unit vector $\ff e_{\ff S} = \ff S / S$ and Eq.\ (\ref{eq:cons}) imply that the ground-state expectation value $\langle \ff s_{i_0} \rangle^{(0)}$ and $\ff S$ must be collinear at any instant of time.
Hence, the linear-in-$J$ torque, i.e., the first term in Eq.\ (\ref{eq:eomberry}) vanishes. 

For the same symmetry reasons we must have $\ff \Omega(\ff S) = \pm \Omega(\ff S) \ff e_{\ff S} = \pm \Omega(S) \ff e_{\ff S}$ where $\Omega \equiv | \ff \Omega | > 0$. 
These considerations lead to an LL-type equation, $\dot{\ff S} = \ff S \times \ff B / (1 \pm S \Omega(S))$, but with a renormalized precession frequency $\omega_{\rm p} = B / (1 \pm S\Omega(S))$, depending on the Berry curvature.

%----------------------------------------------------------------------------------------------------------
\parag{Berry curvature, general considerations.}
 
Integrating the Berry curvature over the surface of the Bloch sphere with radius $S$, we get $\oint \Omega(S) \ff e_{\ff S} d\ff \Gamma = 4 \pi S^{2} \Omega(S)$. 
On the other hand, integration over an arbitrary surface $\Gamma$ yields the gauge invariant Berry phase $\gamma = \int_{\Gamma} \ff \Omega(\ff S) d\ff \Gamma = \int_{\partial \Gamma} \ff C(\ff S) d\ff S$ where $\ff \Omega = \nabla_{\ff S} \times \ff C$ and $\ff C = i \langle \Psi_{0}[\ff S] | \nabla_{\ff S} | \Psi_{0}[\ff S] \rangle$ is the spin Berry connection.
%\cc{note $\Gamma$ has a nontrivial boundary here (in general) and thus Stokes' theorem applies}
For a {\em closed surface}, i.e., a vanishing path $\partial \Gamma$ on the sphere, the Berry phase is quantized (see e.g.\ Refs.\ \cite{TKNN82,NS83}): $\gamma = 2 \pi k$, where the Chern number $k$ is an integer such that $e^{i\gamma}=1$. 

For our case this implies $\Omega(S) = \pm k / 2S^{2}$, and thus the geometrical torque is largely independent of the details of the electronic structure.
This result also provides an interpretation for the Chern number: 
With $\ff S$ interpreted as the position vector, $\ff \Omega(\ff S) = \pm k \ff e_{\ff S} / 2 S^{2}$ is the field strength of a magnetic monopole with magnetic charge $\pm k/2$ at the origin of the Bloch sphere, and $\pm 2\pi k$ is the magnetic flux through the surface \cite{Dir60}.
Note that Stokes' theorem cannot be applied as the corresponding gauge field $\ff C(\ff S)$ is necessarily singular on at least one point on the Bloch sphere. 

%%%%%%%%%%%%%%%%%%%%%%%%%%%%%%%%%%%%%%%%%%%%%%%%%%%%%%%%%%%%%%
\begin{figure}
\includegraphics[width=0.9\columnwidth]{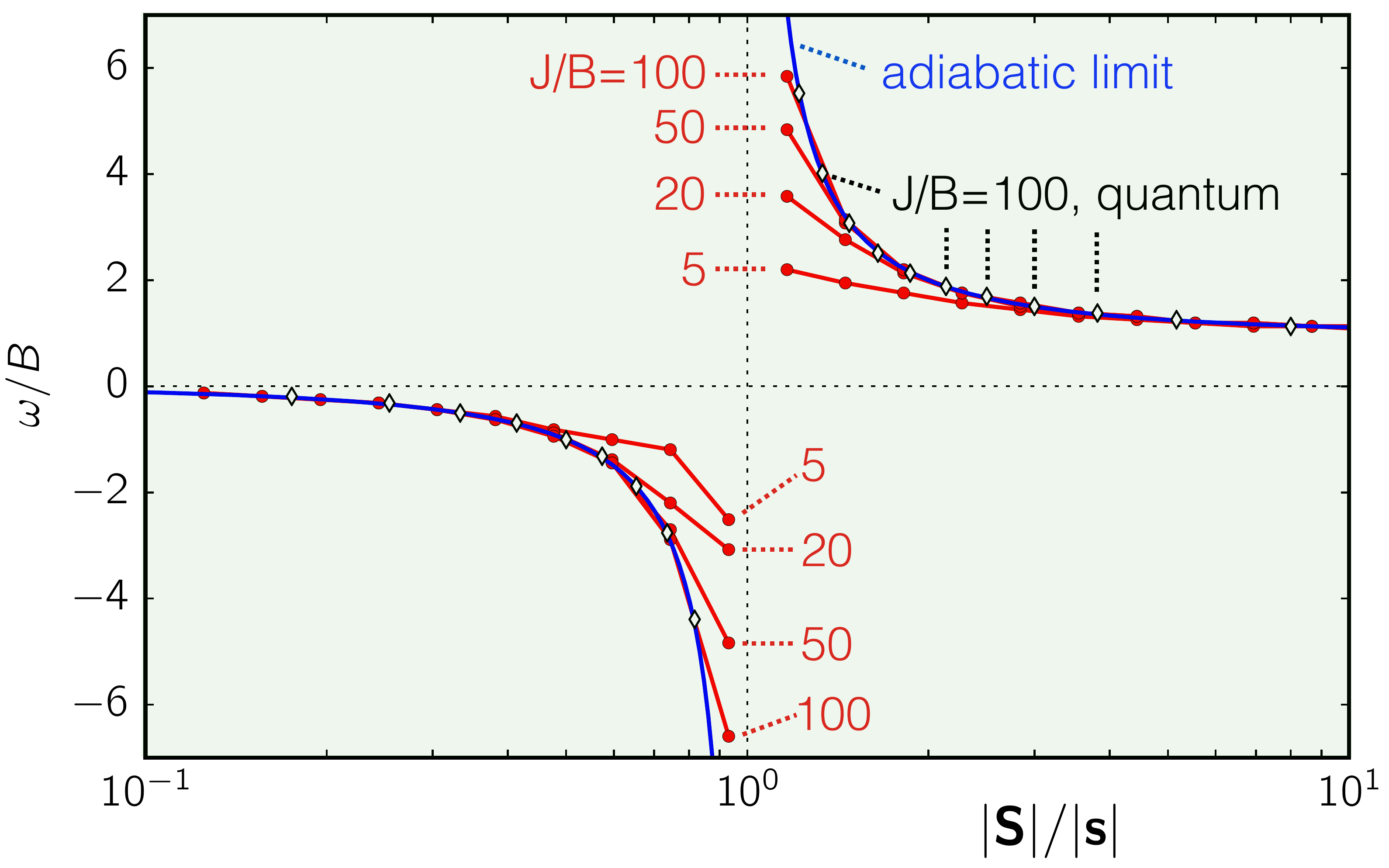}
\caption{
Precession frequency as obtained from adiabatic spin dynamics [Eq.\ (\ref{eq:omega}), blue line] and compared to numerical results obtained for the classical two-spin model at different $J$ (red circles) and for the quantum two-spin model at $J/B=100$ (open diamonds).
}
\label{fig:s1s2}
\end{figure}
%%%%%%%%%%%%%%%%%%%%%%%%%%%%%%%%%%%%%%%%%%%%%%%%%%%%%%%%%%%%%%

%----------------------------------------------------------------------------------------------------------
\parag{Computation of the Berry curvature.}

Since $| \Psi_{0}[\ff S] \rangle = \prod_{n} c_{n}^{\dagger} | \mbox{vac.} \rangle$ must be a Slater determinant for noninteracting electrons, it is straightforward to show that the spin Berry curvature [Eq.\ (\ref{eq:curv})] is additive and can be obtained as a sum over individual contributions $\Omega_{n}$ of the occupied single-particle states $n=(k,\sigma)$ \cite{Res00}.
Furthermore, spin-rotation symmetry requires that the vector $\ff \Omega$ must vanish for any isotropic spin-singlet state, in particular for a Fermi-sea singlet $\prod_{k} c_{k\uparrow}^{\dagger} c_{k\downarrow}^{\dagger} |\mbox{vac.}\rangle$. 
We conclude that $\ff \Omega = (M_{z} s / S^{2}) \ff e_{\ff S}$ is determined by the system's magnetization $M_{z} = N_{\uparrow} - N_{\downarrow}$ times the curvature of a spin-majority one-particle state $n$, which is easily computed as $\Omega_{n} = s / S^{2}$ (with $s=1/2$) \cite{Bru05,XCN10}. 

For the present case of a single classical spin locally coupled to an unpolarized Fermi sea and for antiferromagnetic coupling $J>0$, we have $M_{z} \le 0$. 
Namely, $M_{z} = 0$ for even and $M_{z} = -1/2$ ($k=-1$) for odd particle number $N$, i.e., there is an odd-even effect related to the Kramers degeneracy of the electronic ground state. 
Hence, for odd $N$ we find $\omega_{\rm p} = B / (1 - S \Omega(S)) = B /(1-s/S)$, i.e., Eq.\ (\ref{eq:omega}), while for even $N$ the geometrical torque is absent.

%----------------------------------------------------------------------------------------------------------
\parag{Two-spin model.}

The effects of the geometrical torque manifest themselves in the adiabatic limit.
Most simply, this is realized for $N=L=1$, i.e., effectively for a two-spin model $H_{\rm 2-spin} = J \ff s \ff S - \ff B \ff S$, in the strong-coupling limit $J \gg B$ where $\ff s(t) \equiv \langle \ff s_{i_0} \rangle_{t}$ and $\ff S(t)$ are forced to align antiferromagnetically.
Fig.\ \ref{fig:s1s2} displays the $S$-dependence of the precession frequency obtained numerically for different $J$ (and $s=1/2$) and as predicted by Eq.\ (\ref{eq:omega}). 
An enhanced $\omega_{\rm p} > B$ is found for $S> s$. 
For $S<s$ the geometrical torque even leads to a reversed orientation of the precessional motion.
To ensure that the conduction-electron local moment $\ff s(t)$ follows the spin adiabatically, requires an ever stronger $J/B$ when approaching the limit $S=s=1/2$. 
Stated differently, the adiabatic approximation is never correct for $S=s$. 

%----------------------------------------------------------------------------------------------------------
\parag{Quantum spin.}

Anomalous precession due to the geometrical torque is also found in the quantum variant of $H_{\rm pn}$, i.e., replacing the classical spin by a quantum spin with spin quantum number $S$, see the black open symbols in Fig.\ \ref{fig:s1s2}.
Using the time-dependent density-matrix renormalization group (t-DMRG, see Refs.\ \cite{SRP16b,Sch11,HCO+11}), we have also qualitatively verified the effect for $L>1$ in the quantum-spin case, see Fig.\ \ref{fig:soft} (``quantum''). 
Note that due to the (underscreened) Kondo effect we have $| \langle \ff S \rangle_{t=0} | < 1$ and that due to restrictions of the t-DMRG we are limited to a smaller system ($L=51$) and time scale ($t<100$) in this case.
Still one can read off a precession frequency of $\omega_{\rm p} \approx 1.3B > B$ which is consistent with Eq.\ (\ref{eq:omega}) in the form $\omega_{\rm p} \approx B / (1 - | \langle \ff s_{i_{0}} \rangle | / | \langle \ff S \rangle | )$.

%%%%%%%%%%%%%%%%%%%%%%%%%%%%%%%%%%%%%%%%%%%%%%%%%%%%%%%%%%%%%%
\begin{figure}
\includegraphics[width=0.98\columnwidth]{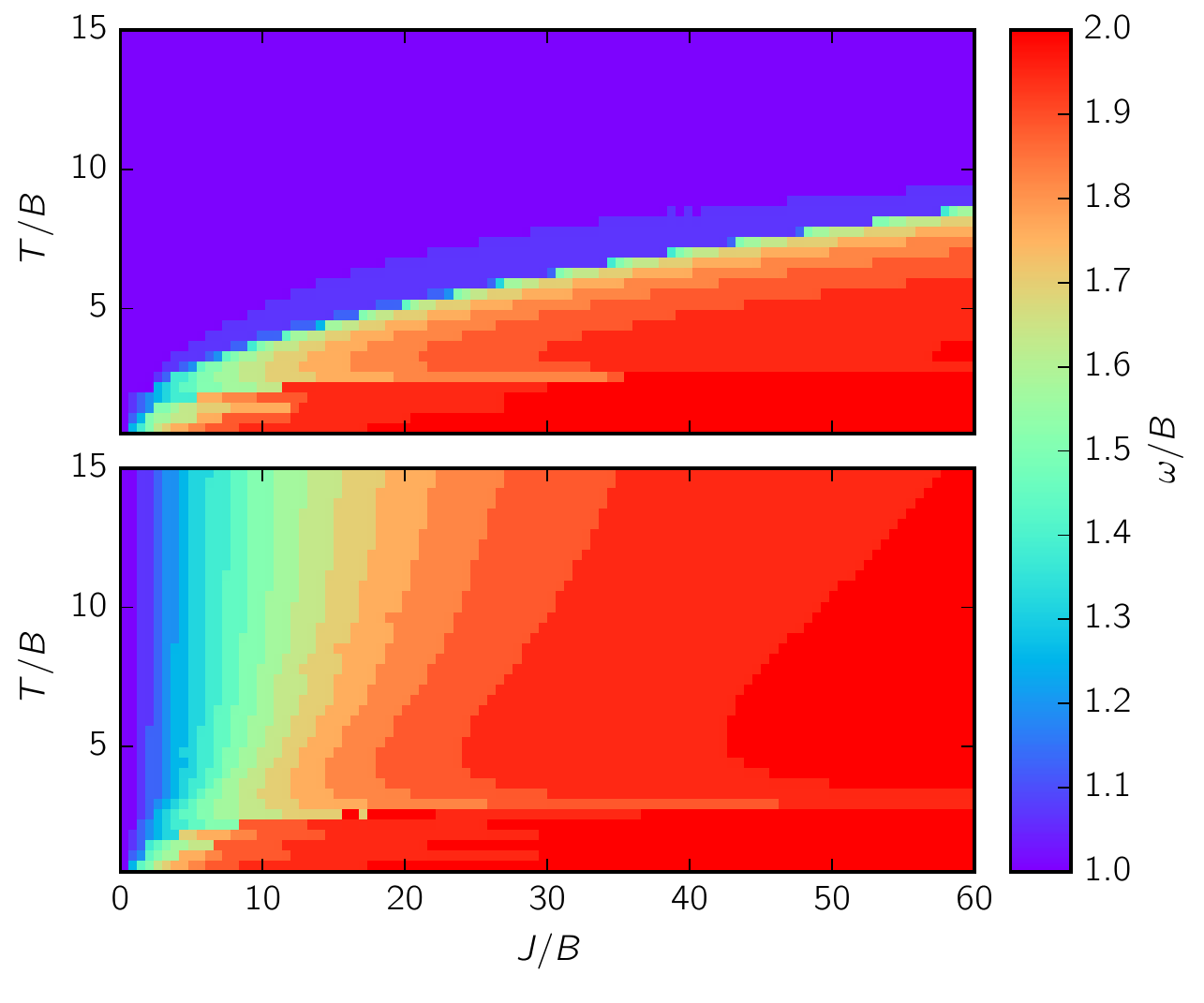}
\caption{
Precession frequency $\omega_{\rm p}/B$ (see color code on the right) as obtained numerically for the $s$-$d$ model (\ref{eq:ham}) with a classical spin $S=1$, at half-filling $n=1$, and for open chains with even ($L=10$, {\em top}) and odd ($L=11$, {\em bottom}) number of electrons $N=L$. 
Results from $1800+1800$ independent calculations for different $T/B$ and $J/B$.
}
\label{fig:om}
\end{figure}
%%%%%%%%%%%%%%%%%%%%%%%%%%%%%%%%%%%%%%%%%%%%%%%%%%%%%%%%%%%%%%

%----------------------------------------------------------------------------------------------------------
\parag{Dynamical phase diagram for $S=1$.}

To distinguish between regimes with normal and anomalous precession frequencies for the quantum-classical model (\ref{eq:ham}), we have numerically solved Eqs.\ (\ref{eq:eoms}) and (\ref{eq:rhoeom}) and computed the precession frequency in a large range of model parameters, see Fig.\ \ref{fig:om}.
We essentially identify two mechanisms leading to adiabatic spin dynamics and therewith to anomalous precession:

First, for $JS > 2T$ the local magnetic field $J\ff S$ creates bound electronic states, localized at $i_{0}=1$ and split off from the conduction band. 
Hence, in the strong-coupling limit $JS \gg T$ the dynamics is effectively captured by the two-spin model, weakly distorted on an energy scale $T^{2}/J$, and thus exhibits an anomalous procession frequency $\omega_{\rm p} \approx 2B$, as predicted by Eq.\ (\ref{eq:omega}), if $J \gg B$. 

This is found in the $J / B > (T/B)^{2} >1$ regime of Fig.\ \ref{fig:om} (upper panel).
Note that effectively only the local particle number $\sum_{\sigma} \langle n_{i_{0}\sigma} \rangle =1$ at $i_{0}$ counts. 
The mechanism is thus independent of the total particle number $N$. 

For odd $N$ (lower panel of Fig.\ \ref{fig:om}) the anomalous regime is more extended. 
This is a consequence of the adiabatic theorem.
Roughly, adiabaticity is ensured \cite{Com09,AE99} if the variation rate of the {\em electronic} Hamiltonian $\sim B$ is small compared to the finite-size gap $\Delta$. 
For weak $J < T$ and for odd $N$ (to get a finite Berry curvature), the {\em local} perturbation $J\ff S \ff s_{i_{0}}$ induces a spin-splitting $\Delta \sim JS / L + \ca O(J^{2})$.
Hence, we need $B \ll JS / L$; see the $J/B > L = 11$ regime in the lower panel (odd $N$) of Fig.\ \ref{fig:om}.
This appears realistic as it requires the ability to control the magnetic degrees of freedom on a time scale much {\em larger} than $L/J$ which, for nano systems (say, $L=100$) and for $J \sim 1 \mbox{meV}$, is in the subnanosecond regime.

%--------------------------------------------------------------------------------------------------------------
\parag{Conclusions and outlook.}

Since the time scales for spin and electron dynamics are typically well separated, an adiabatic spin-dynamics approach is highly interesting conceptually.
In addition, as an important practical advantage, the resulting effective spin-only theory does not suffer from the severe restrictions on the accessible time and length scales inherent to the complete numerical solution of the equations of motion (\ref{eq:eoms}) and (\ref{eq:rhoeom}) of the full theory.
The present study has shown that geometrical torques naturally enter the stage in this context and may dramatically change the fundamental precession frequency. 
This must be expected to have substantial and yet unexplored consequences also for the case of multi-impurity or dense lattice models. 
An according extension of the present theory is straightforward.

We expect that geometrical torques play an important role in spin dynamics generally. 
In particular, the focus on nano systems, adopted here to apply the adiabatic theorem, can be relaxed:
On time scales much longer than the scales set by relaxation mechanisms involving different (e.g.\ phonon) dissipation channels, and at low temperatures, a nearly adiabatic evolution of the electronic system is at least highly plausible.

Since control is required on long time scales only, a direct experimental observation of anomalous precession is well conceivable.
It appears attractive, for example, to envisage $\omega_{\rm p}$ as a ``sensor'' for the Berry curvature, in particular for strongly correlated electron systems, for noncollinear magnetic systems, for topological insulators, or for systems where the electronic degrees of freedom are captured by localized (classical or quantum) spins.

%--------------------------------------------------------------------------------------------------------------
\parag{Acknowledgement.}
We thank Antonia Hintze, Thore Posske and Roman Rausch for stimulating discussions.
This work has been supported by the Deutsche Forschungsgemeinschaft through the SFB 668 (project B3), through the SFB 925 (project B5) and through the excellence cluster ``The Hamburg Centre for Ultrafast Imaging - Structure, Dynamics and Control of Matter at the Atomic Scale''.

\end{document}